\documentclass[dvips]{article}
\usepackage{supertabular,lscape,epsfig}
\usepackage{amssymb}
\usepackage{amsmath}
\DeclareSymbolFont{ppa}{OT1}{ppl}{m}{it}
\DeclareMathSymbol{\vv}{\mathalpha}{ppa}{'166}

\DeclareSymbolFont{ppa}{OT1}{ppl}{m}{it}
\DeclareMathSymbol{\vv}{\mathalpha}{ppa}{'166}

\begin{document}

\newcommand{\dd}{\,{\rm d}}
\newcommand{\ie}{{\it i.e.},\,}
\newcommand{\etal}{{\it et al.\ }}
\newcommand{\eg}{{\it e.g.},\,}
\newcommand{\cf}{{\it cf.\ }}
\newcommand{\vs}{{\it vs.\ }}
\newcommand{\zdot}{\makebox[0pt][l]{.}}
\newcommand{\up}[1]{\ifmmode^{\rm #1}\else$^{\rm #1}$\fi}
\newcommand{\dn}[1]{\ifmmode_{\rm #1}\else$_{\rm #1}$\fi}
\newcommand{\upd}{\up{d}}
\newcommand{\uph}{\up{h}}
\newcommand{\upm}{\up{m}}
\newcommand{\ups}{\up{s}}
\newcommand{\arcd}{\ifmmode^{\circ}\else$^{\circ}$\fi}
\newcommand{\arcm}{\ifmmode{'}\else$'$\fi}
\newcommand{\arcs}{\ifmmode{''}\else$''$\fi}
\newcommand{\MS}{{\rm M}\ifmmode_{\odot}\else$_{\odot}$\fi}
\newcommand{\RS}{{\rm R}\ifmmode_{\odot}\else$_{\odot}$\fi}
\newcommand{\LS}{{\rm L}\ifmmode_{\odot}\else$_{\odot}$\fi}

\newcommand{\Abstract}[2]{{\footnotesize\begin{center}ABSTRACT\end{center}
\vspace{1mm}\par#1\par
\noindent
{~}{\it #2}}}

\newcommand{\TabCap}[2]{\begin{center}\parbox[t]{#1}{\begin{center}
  \small {\spaceskip 2pt plus 1pt minus 1pt T a b l e}   
  \refstepcounter{table}\thetable \\[2mm]
  \footnotesize #2 \end{center}}\end{center}}

\newcommand{\TableSep}[2]{\begin{table}[p]\vspace{#1}
\TabCap{#2}\end{table}}

\newcommand{\FigCap}[1]{\footnotesize\par\noindent Fig.\  %
  \refstepcounter{figure}\thefigure. #1\par}

\newcommand{\TableFont}{\footnotesize}
\newcommand{\TableFontIt}{\ttit}
\newcommand{\SetTableFont}[1]{\renewcommand{\TableFont}{#1}}

\newcommand{\MakeTable}[4]{\begin{table}[t]\TabCap{#2}{#3}
  \begin{center} \TableFont \begin{tabular}{#1} #4 
  \end{tabular}\end{center}\end{table}}

\newcommand{\MakeTableSep}[4]{\begin{table}[p]\TabCap{#2}{#3}
  \begin{center} \TableFont \begin{tabular}{#1} #4
  \end{tabular}\end{center}\end{table}}
\newcommand{\TabCapp}[2]{\begin{center}\parbox[t]{#1}{\centerline{
  \small {\spaceskip 2pt plus 1pt minus 1pt T a b l e}
  \refstepcounter{table}\thetable}
  \vskip2mm
  \centerline{\footnotesize #2}}
  \vskip3mm
\end{center}}

\newcommand{\MakeTableSepp}[4]{\begin{table}[p]\TabCapp{#2}{#3}\vspace*{-.7cm}
  \begin{center} \TableFont \begin{tabular}{#1} #4 
  \end{tabular}\end{center}\end{table}}

\newfont{\bb}{ptmbi8t at 12pt}
\newfont{\bbb}{cmbxti10}
\newfont{\bbbb}{cmbxti10 at 9pt}
\newcommand{\uprule}{\rule{0pt}{2.5ex}}
\newcommand{\douprule}{\rule[-2ex]{0pt}{4.5ex}}
\newcommand{\dorule}{\rule[-2ex]{0pt}{2ex}}
\def\thefootnote{\fnsymbol{footnote}}

\newenvironment{references}%
{
\footnotesize \frenchspacing
\renewcommand{\thesection}{}
\renewcommand{\in}{{\rm in }}
\renewcommand{\AA}{Astron.\ Astrophys.}
\newcommand{\AAS}{Astron.~Astrophys.~Suppl.~Ser.}
\newcommand{\ApJ}{Astrophys.\ J.}
\newcommand{\ApJS}{Astrophys.\ J.~Suppl.~Ser.}
\newcommand{\ApJL}{Astrophys.\ J.~Letters}
\newcommand{\AJ}{Astron.\ J.}
\newcommand{\IBVS}{IBVS}
\newcommand{\PASP}{P.A.S.P.}
\newcommand{\Acta}{Acta Astron.}
\newcommand{\MNRAS}{MNRAS}
\renewcommand{\and}{{\rm and }}
\section{{\rm REFERENCES}}
\sloppy \hyphenpenalty10000
\begin{list}{}{\leftmargin1cm\listparindent-1cm
\itemindent\listparindent\parsep0pt\itemsep0pt}}%
{\end{list}\vspace{2mm}}

\def\TYLDA{~}
\newlength{\DW}
\settowidth{\DW}{0}
\newcommand{\dw}{\hspace{\DW}}

\newcommand{\refitem}[5]{\item[]{#1} #2%
\def\REFARG{#3}\ifx\REFARG\TYLDA\else, {\it#3}\fi
\def\REFARG{#4}\ifx\REFARG\TYLDA\else, {\bf#4}\fi
\def\REFARG{#5}\ifx\REFARG\TYLDA\else, {#5}\fi.}

\newcommand{\Section}[1]{\section{\hskip-6mm.\hskip3mm#1}}
\newcommand{\Subsection}[1]{\subsection{#1}}
\newcommand{\Acknow}[1]{\par\vspace{5mm}{\bf Acknowledgements.} #1}
\pagestyle{myheadings}

\newcommand{\xrule}{\rule{0pt}{2.5ex}}
\newcommand{\xxrule}{\rule[-1.8ex]{0pt}{4.5ex}}
\def\thefootnote{\fnsymbol{footnote}}
\begin{center}
{\Large\bf The Optical Gravitational Lensing Experiment.\\
\vskip3pt
Miras and Semiregular Variables\\
\vskip6pt
in the Large Magellanic Cloud\footnote{Based on observations obtained with the
1.3~m Warsaw telescope at the Las Campanas Observatory of the Carnegie
Institution of Washington.}}
\vskip1.2cm
{\bf I.~~S~o~s~z~y~\'n~s~k~i$^{1,2}$,~~A.~~U~d~a~l~s~k~i$^1$,~~M.~~K~u~b~i~a~k$^1$,\\
M.\,K.~~S~z~y~m~a~{\'n}~s~k~i$^1$,~~G.~~P~i~e~t~r~z~y~\'n~s~k~i$^{1,2}$,~~K.~~\.Z~e~b~r~u~\'n$^1$,\\
O.~~S~z~e~w~c~z~y~k$^1$,~~\L.~~W~y~r~z~y~k~o~w~s~k~i$^1$,~~and~~K.~~U~l~a~c~z~y~k$^1$}
\vskip8mm
{$^1$Warsaw University Observatory, Al.~Ujazdowskie~4, 00-478~Warszawa, Poland\\
e-mail: \small{(soszynsk,udalski,mk,msz,pietrzyn,zebrun,szewczyk,wyrzykow,kulaczyk)@astrouw.edu.pl}\\
$^2$ Universidad de Concepci{\'o}n, Departamento de Fisica, Casilla 160--C, Concepci{\'o}n, Chile}
\end{center}

\vskip1.6cm

\Abstract{We use the OGLE-II and OGLE-III data in conjunction with the
2MASS near-infrared (NIR) photometry to identify and study Miras and
Semiregular Variables (SRVs) in the Large Magellanic Cloud. We found in
total 3221 variables of both types, populating two of the series of NIR
period--luminosity ({\it PL}) sequences. The majority of these objects are
double periodic pulsators, with periods belonging to both {\it PL}
ridges. We indicate that in the period -- Wesenheit index plane the
oxygen-rich and carbon-rich AGB stars from the NIR {\it PL} sequences C,
C$'$ and D split into well separated ridges. Thus, we discover an effective
method of distinguishing between O-rich and C-rich Miras, SRVs and stars
with Long Secondary Periods using their $V$ and {\it I}-band photometry. We
present an empirical method of estimating the mean $K_s$ magnitudes of the
Long Period Variables using single-epoch $K_s$ measurements and complete
light curves in the {\it I}-band. We utilize these corrected magnitudes to
show that the O-rich and C-rich Miras and SRVs follow somewhat different
$K_s$-band {\it PL} relations.}{Stars: AGB and post-AGB -- Stars: late-type
-- Stars: oscillations -- Magellanic Clouds}

\Section{Introduction}
Our knowledge of the Long Period Variables (LPVs) significantly increased
when large microlensing surveys collected large enough amount of
photometric data to reliably determine the periods and other parameters of
these stars. Wood \etal (1999) showed five parallel sequences in the
period--luminosity ({\it PL}) plane (labeled A--E), each populated by LPVs of
different features. As a luminosity in the {\it PL} diagram Wood \etal (1999)
used reddening free Wesenheit index, but very similar picture was shown for
the near-infrared (NIR) $K$ waveband (Wood 2000). Earlier two of these
sequences were known: {\it PL} relation for Miras (Glass and Lloyd Evans 1981)
and the additional sequence for Semiregular Variables (SRVs, Wood and Sebo
1996).

The subsequent papers confirmed and extended these results (Cioni \etal
2001, Noda \etal 2002, Lebzelter \etal 2002, Cioni \etal 2003). Important
progress in this field has been made due to the analysis of the photometric
data originated in the Optical Gravitational Lensing Experiment (OGLE)
conjuncted with the NIR measurements from various sources (2MASS, SIRIUS,
DENIS). The OGLE project is a large scale photometric survey regularly
monitoring the densest regions of the sky. Collected long-term photometric
data of millions of objects are an ideal material for studying a wide
variety of variable stars, also red giants.

Kiss and Bedding (2003) used OGLE and 2MASS data to show that the Wood's
sequence B above the Tip of the Red Giant Branch (TRGB) is made up with two
sequences. Moreover, below the TRGB there are three sequences which are
shifted in $\log{P}$ relative to the {\it PL} relations above the TRGB. This
was the final proof that the majority of the pulsating variables below the
TRGB are the first ascent giants (RGB stars). These results were confirmed
by Ita \etal (2004), who labeled by C$'$ the new discovered sequence between
the sequences B and C. We adopt this notation in this paper.

Soszy{\'n}ski \etal (2004a) showed the complex structure of the {\it PL}
distribution analyzing the OGLE Small Amplitude Red Giants (OSARGs) --
stars which most frequently can be found in the sequences A and B. When the
secondary periodicities of these multi-periodic variables are taken into
consideration, one can find that these objects lie in the four relatively
narrow {\it PL} sequences. Moreover, Soszy{\'n}ski \etal (2004a)
empirically showed that the variables fainter than the TRGB are a mixture
of the AGB and RGB stars and both groups follow different {\it PL}
relations. Finally, to complicate this pattern, it was shown that the
ridge A below the TRGB consist of three closely spaced parallel sequences,
what is distinctly visible in the Petersen diagram of these variables.

The aim of the present study is to show that the structure of the sequences
C$'$, C and D is also more complex than it looked at the first glance. We
discovered that each of these sequences splits into two ridges in the
period -- optical Wesenheit index plane. These division corresponds to the
spectral separation into oxygen-rich and carbon-rich AGB stars. We use
this feature to separate both spectral types of variables, and show that
O-rich and C-rich Miras and SRVs follow somewhat different {\it PL}
relations in the $K_s$ waveband.

The paper is organized as follows. In Section~2 we describe the $I$ and
{\it V}-band observations and the cross-correlation of the optical and NIR
data. In Section~3 we present the selection of the Miras and SRVs in our
data. Section~4 describes the sample and some features of the {\it PL}
relations are presented. In Section~5 we show period -- optical Wesenheit
index diagram on which the O- and C-rich AGB stars are well
separated. Section~6 contains the description of the mean $K_s$-band
magnitudes estimation. In Section~7 we present {\it PL(K)} diagram for
Miras and SRVs. Our results are discussed and summarized in Sections~8
and~9.

\Section{Observations and Data Reductions}
The optical data used in this analysis were obtained from the 1.3~m Warsaw
Telescope at the Las Campanas Observatory, Chile, operated by the Carnegie
Institution of Washington. The {\it I}-band data span a time baseline of about
3000~days: from January 1997 to April 2005. During the years 1997--2000,
when the second phase of the OGLE survey (OGLE-II) was conducted, the
telescope was equipped with the ``first generation'' camera with the SITe
${2048\times2048}$ CCD detector. The pixel size was 24~$\mu$m what
corresponded to 0.417~arcsec/pixel scale. Observations of the LMC were
performed in the ``slow'' reading mode of the CCD detector with the gain
3.8~e$^-$/ADU and readout noise of about 5.4~e$^-$. Details of the
instrumentation setup can be found in Udalski, Kubiak and Szyma{\'n}ski
(1997).

Since 2001, when the third stage of the project (OGLE-III) started, the
telescope has been equipped with a ``second generation'' eight chip
${8192\times8192}$ pixel CCD mosaic camera (Udalski 2003). The pixel size
of the detector is 15~$\mu$m, giving the 0.26 arcsec/pixel scale and the
field of view about ${35\arcm\times35\arcm}$. The gain of each chip is
adjusted to be about 1.3~e$^-$/ADU with the readout noise from 6 to 9~e$^-$
depending on the chip.

{\it V}-band measurements were collected during the second phase of the
OGLE experiment, so they span shorter time baseline. Up to 70 {\it V}-band
points per star is available. In the {\it I}-band we collected in total from
500 to 900 measurements, depending on the field. The observations were
reduced with the Difference Image Analysis developed by Alard and Lupton
(1998) and Alard (2000), implemented by Wo{\'z}niak (2000) and Udalski
(2003). The transformations of the instrumental photometry to the standard
system, and the determination of the equatorial coordinates of stars is
described by Udalski \etal (2000).

The NIR $K_s$ and $J$ measurements used in this work come from the 2MASS
All-Sky Catalog of Point Sources (Cutri \etal 2003). We performed the
spatial cross-correlation of the OGLE variables with the 2MASS catalog in
two steps. In the first step for each OGLE star we found the closest object
in the 2MASS list. Then, for each OGLE field we determined the mean offset
between OGLE and 2MASS positions. The typical value of the offset was equal
to 0.5 arcsec in RA$\cos(\delta)$ and 0.1 arcsec in DEC. In the second step
of the procedure we shifted the 2MASS positions by the offsets and repeated
the cross-identification of objects using the 1 arcsec search radius. That
way we successfully identified more than 98\% of the OGLE red stars
($V-I>0.5$~mag) brighter than $I=17$~mag.

For further analysis we included only sources detected in four photometric
band-passes: $V$, $I$, $J$ and $K_s$, and stars with $V-I>0.5$~mag. The data
have been dereddened using the extinction maps provided by Udalski \etal
(1999) with the extinction law by Schlegel \etal (1998).

\Section{Selection of the Sample}
We performed a period search in the same manner as described by
Soszy{\'n}ski \etal (2004a). Briefly, we used period-searching program {\sc
Fnpeaks} (Ko{\l}aczkowski 2003, private communication) which implements the
algorithm of Discrete Fourier Transform. We tested periodicity of every
{\it I}-band light curve brighter than 17~mag. After finding the highest
peak in the power spectrum, we fitted the third order Fourier series to the
folded light curve and subtracted the function from the data. Then, we
repeated the procedure using the residuals until we found five periods per
star. During the search we determined the {\it I}-band amplitudes for each
period by deriving the difference between the maximum and minimum values of
the fitted function.

We identified Miras and SRVs in the LMC using two features: {\it I}-band
amplitudes and position in the period -- NIR Wesenheit index diagram. The
Wesenheit index (Madore 1982) is a reddening-free quantity defined as a
linear combination of the selected luminosity and color of the star:
$$W_{\lambda_1\lambda_2}=m_{\lambda_2}-R_{\lambda_1\lambda_2}
(m_{\lambda_1}-m_{\lambda_2})$$
where $R_{\lambda_1\lambda_2}$ is the ratio of total-to-selective absorption
at given wavebands:
$$R_{\lambda_1\lambda_2}=\frac{A_{\lambda_2}}{E(\lambda_1-\lambda_2)}=
\frac{R_{\lambda_2}}{R_{\lambda_2}-R_{\lambda_1}},$$
$$R_{\lambda}=\frac{A_{\lambda}}{E(B-V)}.$$

Usually the optical magnitudes and colors are utilized to construct this
quantity, for example:
$$W_{VI}\equiv W_I=I-1.55(V-I).$$

However, the Wesenheit index can be defined also for the NIR data, for
example $K_s$-band magnitudes and $(J-K_s)$ colors. Having Schlegel's 
\etal (1998) ratios of total-to-selective absorption ($R_J=0.902$,
$R_K=0.367$) we obtained the following definition of the NIR Wesenheit
index:
$$W_{JK}=K_s-0.686(J-K_s).$$

\begin{figure}[t]
\centerline{\includegraphics[width=13.91cm]{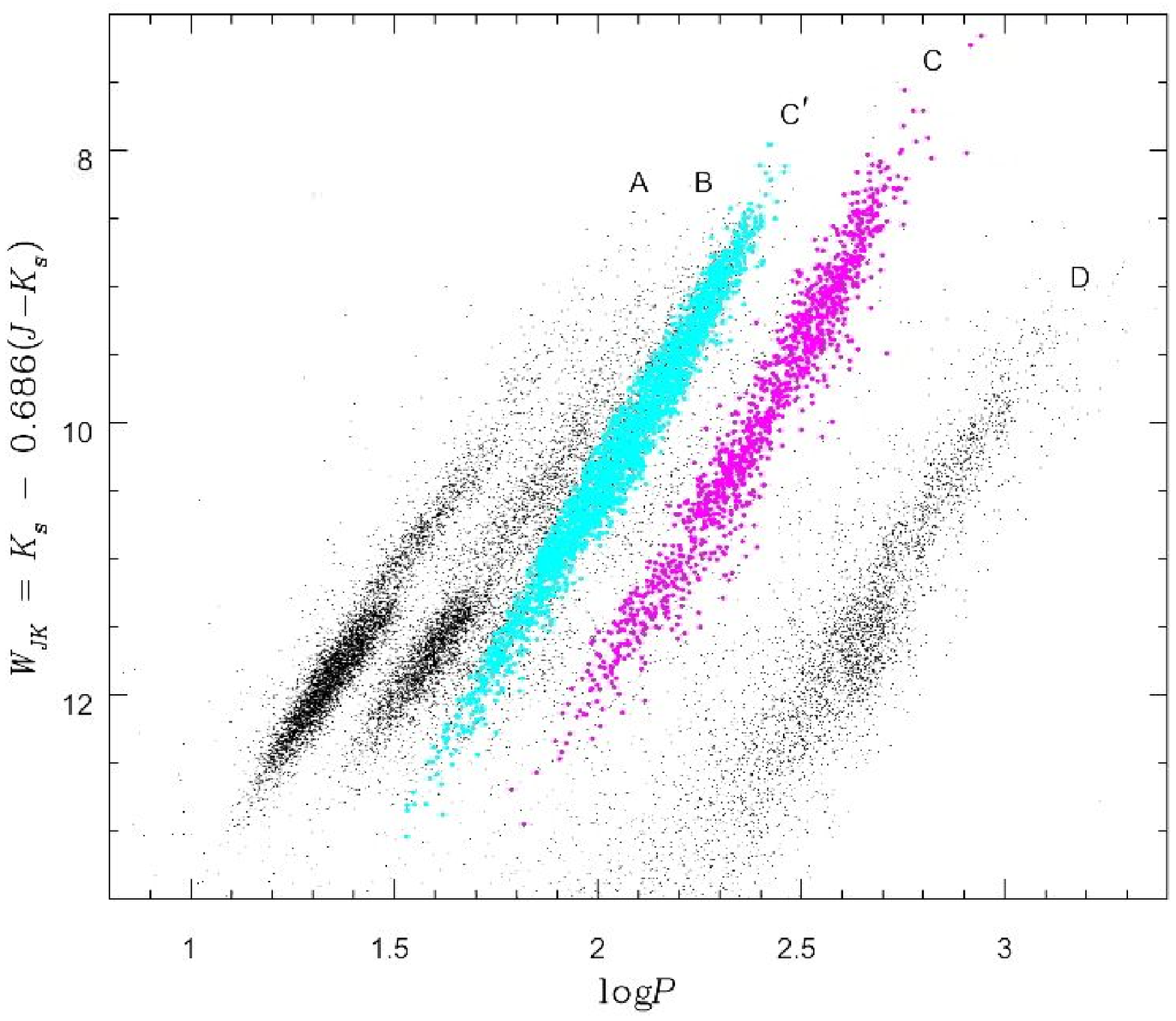}} 
\FigCap{Period--$W_{JK}$ diagram for Miras and SRVs in the LMC. The sequences
C$'$ and C are highlighted in cyan and magenta, respectively.}
\end{figure} 

Fig.~1 shows the period--$W_{JK}$ diagram for red giants in the LMC. The
{\it PL} sequences are better defined than in the $K_s$-band {\it PL}
diagram, in particular when the random-phase single-epoch measurements are
plotted. We used amplitudes and a position in the $\log{P}$--$W_{JK}$
diagram to identify our sample of Miras and SRVs.

Before selecting the stars populating the sequence C we excluded variables
with amplitudes smaller than 0.1~mag. In the case of the sequence C$'$ we cut
the sample at $A(I)=0.03$~mag. We checked that below these limits these
sequences are not visible among real or artificial small amplitude
variables (see also Groenewegen 2004).

Then, we selected LPVs from the sequences C and C$'$ using their position in
the $\log{P}$--$W_{JK}$ diagram (Fig.~1). Because each star in Fig.~1 is
represented by a single-epoch observation in $K_s$ and {\it J}-bands, we
employed the rule that the larger the amplitude of variability, the larger
possible deviation from the mean {\it PL} law. In the final step of our
procedure all selected light curves were carefully checked
visually. Obvious eclipsing and ellipsoidal variables, foreground high
proper motion stars, artifacts etc. were removed from the sample.

It is worth noting that an additional dim sequence is visible between
sequences C$'$ and C. It is built by rather small amplitude variables
($0.02~{\rm mag}<A(I)<0.2$~mag). The same sequence is noticeable in Fraser
\etal (2005) (their Fig.~1).

\Section{Miras and Semiregular Variables}
We selected 1112 variables with the primary period in the sequence C and
2109 objects populating sequence C$'$. This division is in many cases
ambiguous, because multiperiodicity is a common feature among these
stars. About half of the stars populating the sequence C and about 70\% of
objects with dominant periods in the sequence C$'$ have one of the secondary
periods obeying the other {\it PL} relation. Sometimes the amplitudes of
both modes are similar, and the determination of the main periods could be
different, if somewhat different set of data is used.

The list of all variables is available in the electronic form from the 
OGLE {\sc Internet} archive:
\begin{center}
{\it http://ogle.astrouw.edu.pl/} \\
{\it ftp://ftp.astrouw.edu.pl/ogle/ogle2/var\_stars/lmc/lpv/}\\
\end{center}
or its US mirror
\begin{center}
{\it http://bulge.princeton.edu/\~{}ogle/}\\
{\it ftp://bulge.princeton.edu/ogle/ogle2/var\_stars/lmc/lpv/}\\
\end{center}

We provide the equatorial coordinates, primary periods, $V$, $I$, $J$ and
$K_s$ magnitudes, amplitudes and classification of the stars. Individual
{\it BVI} measurements of all objects and finding charts are also available
from the OGLE {\sc Internet} archive. The lists contain together 3586
entries but only 3221 objects, because 365 stars were detected twice -- in
the overlapping regions of adjacent fields.

The sequences C and C$'$ spread down to the luminosities more than 1~mag
below the TRGB. However, we do not find any bump in the luminosity function
below the TRGB, as it was visible for the OSARG variables (Ita
\etal 2002, Kiss and Bedding 2003). This likely means that the sequences
C and C$'$ are composed solely with the AGB stars.

Practically all Miras, \ie variables with $A(V)>2.5$~mag (Kholopov \etal
1985) or $A(I)>0.9$~mag, fall on the sequence C. The remaining stars in
this ridge are formally SRVs, because their amplitudes do not fulfill these
arbitrary criteria. Nevertheless, we did not find any distinct boundary
between Miras and SRVs and the sequence C, what suggests that these stars
represent the same type of variables. To the end of this paper we will call
them ``Mira-like variables''.

There are strong arguments that the sequences C and C$'$ represent
fundamental and first overtone modes of pulsation, respectively, although
this question is not settled yet. Wood \etal (1999) suggested that the next
sequences (B and A, \ie the OSARG variables) are populated by the next
successive modes: the second and higher overtones. This idea was shared by
many other authors, although the theoretical {\it PL} relations do not fit
the observed sequences (Ita \etal 2004, Fraser \etal 2005). Soszy{\'n}ski
\etal (2004a) showed that there are important empirical arguments that
OSARGs are different types of pulsating stars, and they are not the higher
overtones of Miras and SRVs. The analysis of the secondary periodicities of
these variables showed that OSARGs populate four narrow {\it PL} sequences,
different than sequences C$'$ and C. Moreover, the {\it PL} sequence A and B
(and other OSARG's sequences) are roughly parallel to the sequences C and
C$'$ in the $\log{P}$--$K$ diagram, but these ridges have different slopes in
the $\log{P}$--$W_I$ plane. The sequence B even crosses the sequence C$'$ in
the period--$W_I$ diagram at brighter magnitudes. Of course, the {\it PL}
relations formed by the successive overtones should have the same relative
positions on each type of the {\it PL} diagram.

\begin{figure}[p]
\centerline{\includegraphics[width=13.9cm]{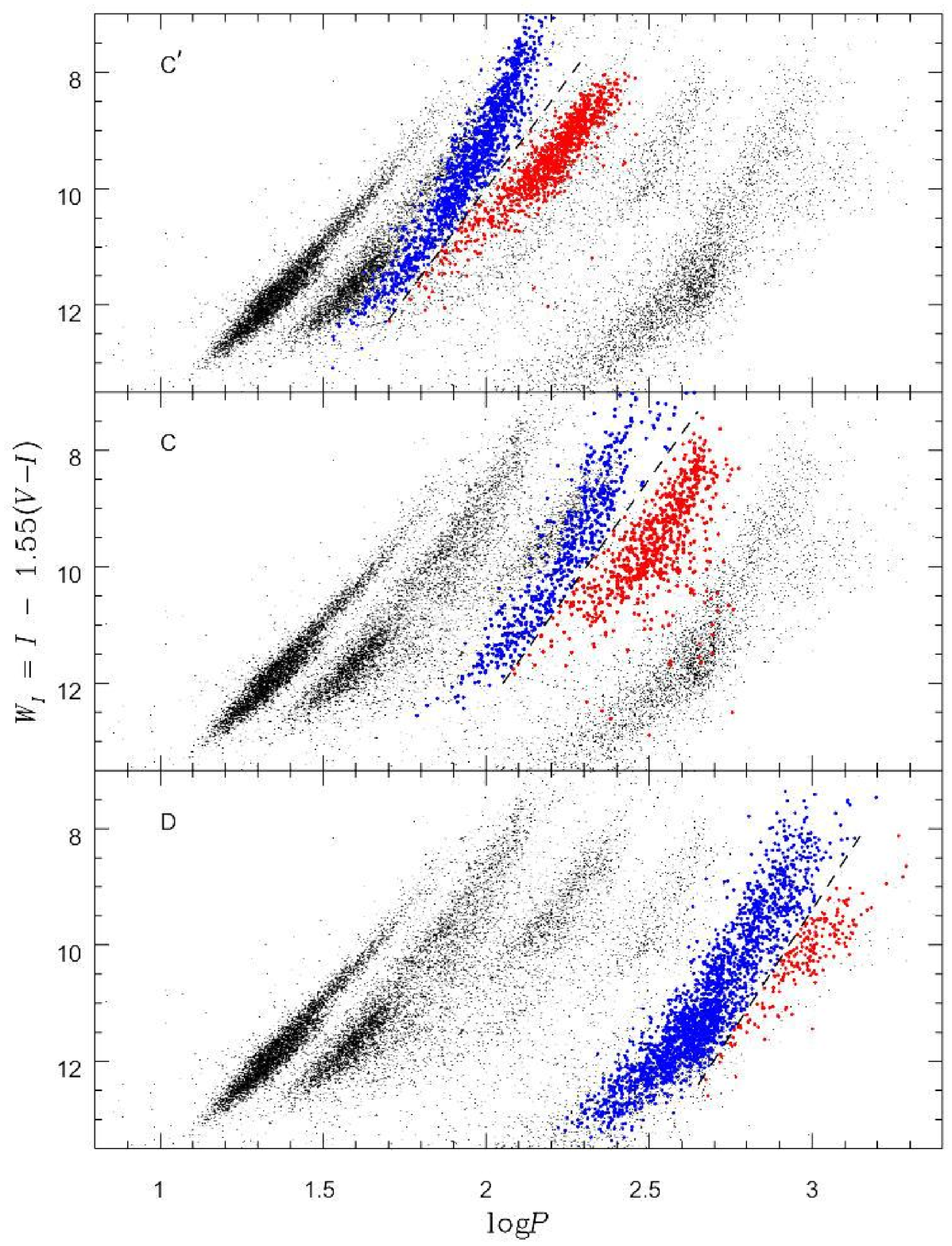}} 
\FigCap{Period--$W_I$ diagrams for LPVs in the LMC. The color points
indicate the stars from the sequences C$'$ ({\it upper panel}), C ({\it
middle panel}) and D ({\it lower panel}). The dashed lines mark the
boundaries selected to separate two distinct groups of objects in each
panel. Different colors refer to different groups of objects.}
\end{figure} 

\begin{figure}[p]
\centerline{\includegraphics[width=13.9cm]{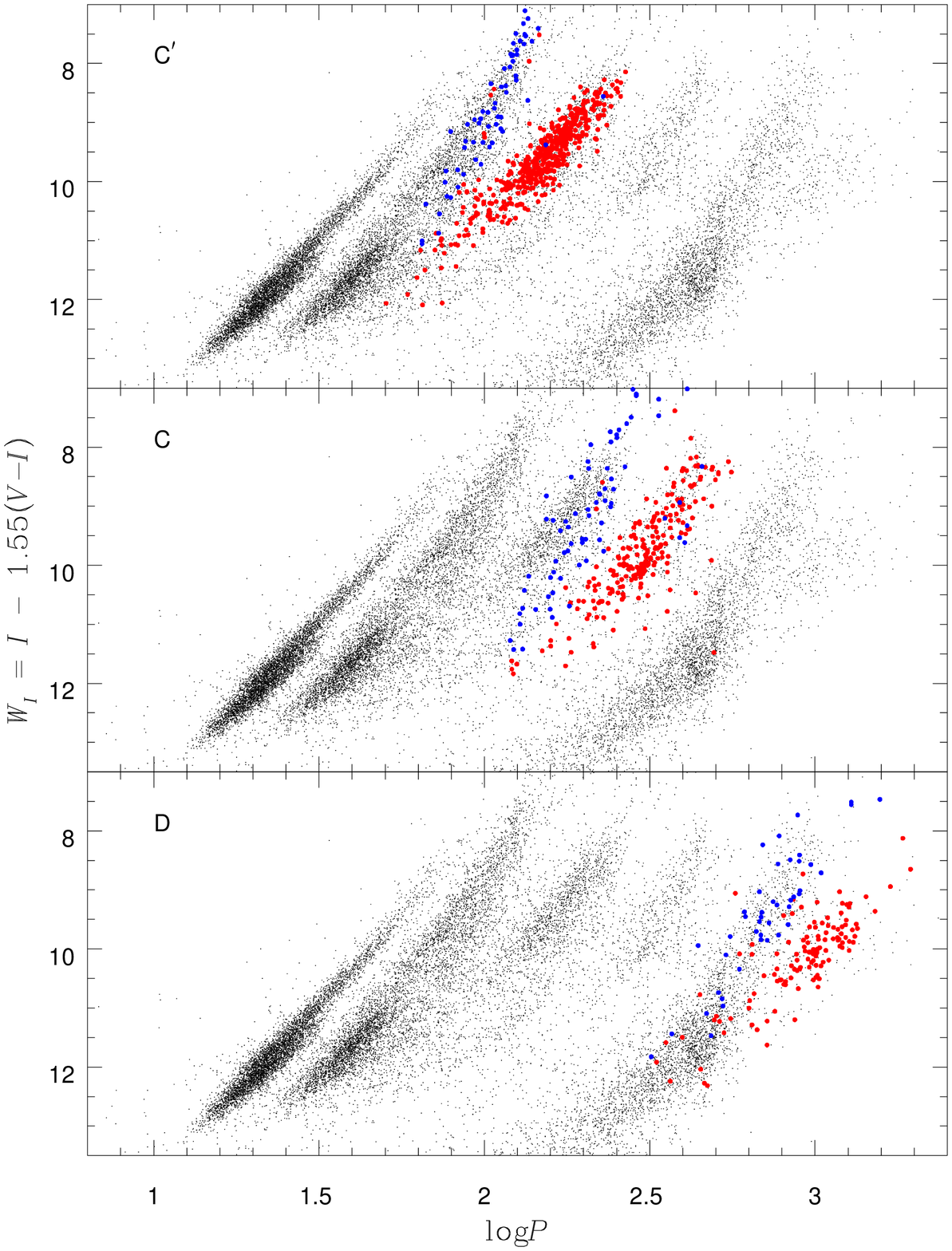}} 
\FigCap{The same as in Fig.~2, but the color points represent only
spectroscopically confirmed O-rich (blue points) and C-rich (red
points) AGB stars (Groenewegen 2004).}
\end{figure} 

\Section{O-rich and C-rich AGB stars}
The AGB stars can be divided into two main spectral classed: oxygen-rich
(O-rich or M-type giants) and carbon-rich (C-rich or C-type), depending on
the relative abundance of oxygen to carbon in their atmospheres. According
to the evolutionary models every star entering the AGB is the O-rich
object. During the evolution the dredge-up episodes change the stellar
surface chemical composition and turn the star from O-rich into C-rich
(Iben and Renzini 1983).

Fig.~2 presents the period--$W_I$ diagrams, separately for stars from the
sequences C$'$, C and D (populated by stars with the Long Secondary
Periods). Since the light curves of many variables in our sample show
secular variability (sometimes with amplitudes of several magnitudes) and
the $I$ and {\it V}-band data span different time baseline, using the mean
$I$ and $V$ luminosities in deriving the $W_I$ index may result in
significant errors. Therefore, we derived individual values of $W_I$ for
the epochs of our {\it V}-band points, estimating the {\it I}-band
magnitudes using the closest measurements. Then, we derived the mean $W_I$
by fitting a third order Fourier series to our individual $W_I$ points and
integrating the function.

The most conspicuous feature visible in Fig.~2 is the fact that each {\it
PL} sequence splits into two ridges in the $\log{P}$--$W_{I}$ plane. The
slanted lines were drawn ``by hand'' to discriminate both groups in each
panel:
$$W_I=-7.5\log{P}+25.0,$$
$$W_I=-7.8\log{P}+28.0,$$
$$W_I=-8.6\log{P}+35.2$$
for the sequences C$'$, C and D, respectively.

Noda \etal (2004) used the photometry of the LMC AGB stars originated in
the MOA project and plotted the period -- (visual) color diagrams
separately for each {\it PL} sequence. They found that the sequences B
(corresponding to the merged sequences B and C$'$ in this paper) and C are
split into two groups in this plane. This dichotomy in Miras and SRVs was
explained by the coexistence of O-rich and C-rich giants.

This hypothesis can be easily checked by plotting the period--$W_I$ diagram
for the spectroscopically confirmed M- and C-type stars. The list of
confirmed objects of both spectral classes was compiled by Groenewegen
(2004), who identified among the OGLE variables in the LMC 1064 C-rich and
344 O-rich AGB stars. The period--$W_I$ diagrams for these objects are
shown in Fig.~3. There is no doubt that the O-rich stars populate the
shorter-period ridges, while the C-rich objects fall in the longer-period
sequences.

There are several stars of both spectral types which do not follow this
rule, \ie some C-rich giants occupy the shorter-period $\log{P}$--$W_{I}$
sequences, and the O-rich giants lie in the longer-period sequences. It can
be explained taking into account that some of the stars have an uncertain
spectral classification. Groenewegen (2004) showed a number of objects with
a different classification in different surveys. He adopted the most recent
determination of the spectral type, but the real classification of these
objects is, in fact, very uncertain.

Hereby, we provide the method of the discriminating the O- and C-rich 
LPVs using their $V$ and $I$ magnitudes. Another commonly used photometric
method of separating these two populations is the discrimination in the
$(J-K_s)$--$K_s$ diagram. O-rich stars are considered to populate colors
$(J-K_s)<1.4$~mag, C-rich stars: $1.4~{\rm mag}<(J-K_s)<2.0$~mag, redder
objects are thought to be the AGB stars of both classes surrounded by thick
dust shells (Nikolaev and Weinberg 2000). Cioni \etal (2005) improved the
discrimination criteria using slanted boundary line in the $(J-K_s)$--$K_s$
diagram.

In Fig.~4 we show such color--magnitude diagrams, separately for the
sequences C and C$'$. Blue dots indicate the O-rich objects, while red
points show the C-rich giants selected in the period--$W_I$ diagram. As can
be expected the M-type objects populate narrow range of $(J-K_s)$ colors,
the C-type stars spread over much wider range. However, there is a region
occupied by both classes of stars, so the separation of the O-rich and
C-rich AGB stars in the NIR color--magnitude diagram can be correct only in
the statistical sense, but in the individual cases may be wrong. Our method
of distinguishing O- and C-rich Miras and SRVs seem to be more reliable,
because in the $\log{P}$--$W_I$ plane both groups are better separated,
especially for brighter objects.

\begin{figure}[p]
\centerline{\includegraphics[width=14cm]{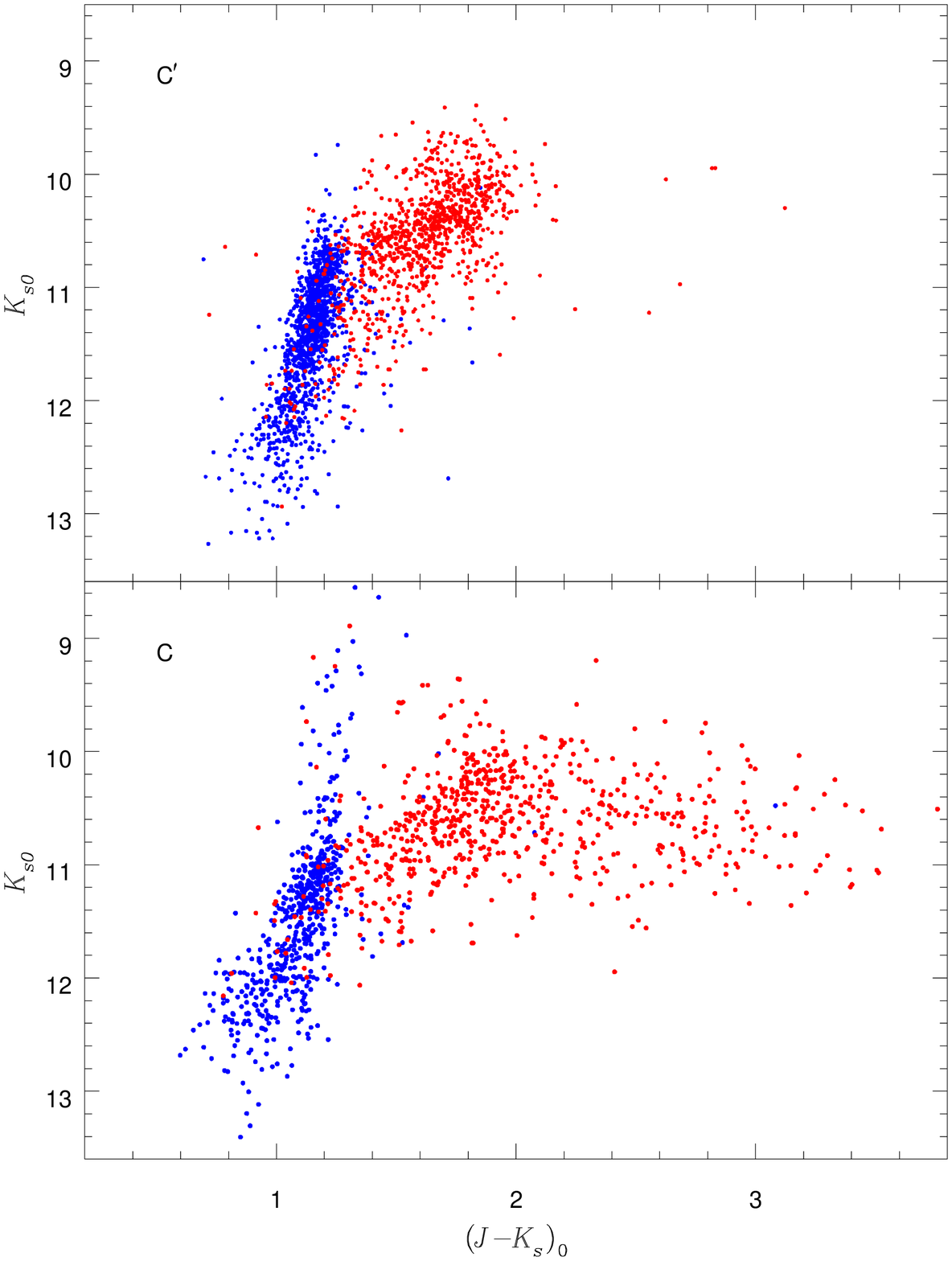}} 
\FigCap{Color--magnitude ($J-K_s,K_s$) diagram for stars populating the
sequences C$'$ ({\it upper panel}) and C ({\it lower panel}). Blue points
represent the O-rich stars, while the red points indicate the C-rich
giants.}
\end{figure}

\Section{Mean ${\pmb K}_{\pmb s}$-band Magnitudes}
The internal scatter of the {\it PL} sequences in the $K_s$-band is
affected by many factors. When only one random-phase observation per star
is available, the {\it PL} relationship is significantly widened by the
amplitude of variability. This effect is particularly noticeable in the
sequence C populated by the stars with the largest amplitudes. According to
Feast \etal (1982) the upper limit for the amplitudes of Miras in the $K$
bandpass is about 1~mag. Thus, the offset in the $K$ magnitude in respect
to the mean {\it PL} relationship for Miras can be as large as 0.5~mag.

To decrease this effect we tried to find a method of converting single
epoch $K_s$-band measurements to the mean magnitudes using the complete
{\it I}-band light curves. Our estimation is based on shifting phases of
the {\it I}-band light curves by a constant offset and scaling their
amplitudes. This approach assumes that the shapes of the light curves in
different wavebands differ only in amplitudes and phases, what is not
true. It is known that Miras in the $K_s$-band have much more sinusoidal
light curves than in the optical part of the spectrum. Thus, our algorithm
can recover the mean magnitudes very roughly, but enough to significantly
decrease the dispersion of the {\it PL} relations.

Because many of the Mira-like variables show secular or very long period
variations, we determined the {\it I}-band luminosity using unfolded light
curves. Whitelock \etal (2003) noticed that the amplitudes of these
long-term trends depend on the wavelength in roughly the same way as the
pulsation amplitudes, \ie they are smaller at longer wavelengths. For the
Julian Date of the $K_s$ observation we added the proper phase lags (see
below) and approximated the {\it I}-band luminosity for this epoch using
three closest points of the light curve. Then, the difference between the
obtained luminosity and the mean {\it I}-band magnitude were scaled using
the $A(K_s)/A(I)$ amplitude ratios, and this value was subtracted from the
$K_s$-band measurement to obtain the mean $K_s$-band magnitude.

It is known from decades that the NIR maxima of Miras occur with a phase
lag of about 0.1 period with respect to the visual light curves (Pettit and
Nicholson 1933), but this phenomenon is still rather poorly
documented. Recently Smith \etal (2005) studied the optical--NIR phase
shifts in a number of AGB stars and confirmed that there is a lag between
the visual and NIR light curves for Miras, while most of the SRVs do not
reveal any detectable phase lags. It is also known that the amplitudes of
the giant variability depend on the photometric bandpass, \ie the
amplitudes decrease with the wavelength. Lebzelter and Wood (2005) found
the ratio of the $K$ and $V$ amplitudes to be equal to about 0.2.

Since the dependence of the NIR--visual amplitudes and phases is rather
poorly known, we decided to estimate the typical amplitude ratios and lag
in phases between the {\it I}-band and $K_s$-band light curves for our
sample of Miras and SRVs. We used two methods. In the first approach we used
two independent $K_s$-band measurements for each star: one coming from the
2MASS catalog and the other from the 3rd release of the DENIS catalog. The
cross-identification of our objects with the DENIS sources was performed in
the same manner as in the case of the 2MASS catalog (Section~2). The DENIS
$K_s$-band magnitudes were transformed to the 2MASS system using the
equations provided by Carpenter (2001). Then, we tested a range of the
phase shifts ($0.0<\phi_K-\phi_I<0.2$) and amplitude ratios
($0.1<A(K_s)/A(I)<0.6$), deriving for each pair of parameters the mean
$K_s$ magnitudes independently for the 2MASS and DENIS single epoch
measurements. For each star we calculated the difference between the mean
luminosities estimated from these two sources, and determined the
dispersion of these differences for the whole sample of stars. Finally, we
selected the phase offset and the amplitude ratio providing the best
agreement between the mean magnitudes estimated using the 2MASS and DENIS
measurements.

In the second method we used the 2MASS data only. We chose the
$\phi_K-\phi_I$ and $A(K_s)/A(I)$ for which the $K_s$-band {\it PL}
relation had the smallest dispersion. We applied both methods separately
for four groups of variables: O- and C-Mira-like stars and O- and C-rich
objects from the sequence C$'$.

The results of these methods appeared to be consistent, and they agree with
the previous results. The bulk of both spectral types of stars located in
the sequence C reveal the same values of optical--NIR phase lags and
amplitude ratios: $\phi_K-\phi_I\approx0.1$, $A(K_s)/A(I)\approx0.4$. 
However, the possible values of the phase shifts in the C-rich Miras seem
to cover wider range, from 0.05 to 0.1. The Miras' amplitude ratios agree
with the results of Lebzelter and Wood (2005), because the $A(I)/A(V)$
ratio is on average 0.5. For the stars populating the sequence C$'$ the
minimum dispersion shows the lag in phase by about 0.02, but probably in
some cases it is somewhat larger, up to 0.05. We found different amplitude
ratios for different spectral types: $A(K_s)/A(I)\approx0.25$ for the
O-rich and $A(K_s)/A(I)\approx0.4$ for the C-rich SRVs in the sequence C$'$.

\begin{figure}[t]
\centerline{\includegraphics[width=9.5cm,angle=270]{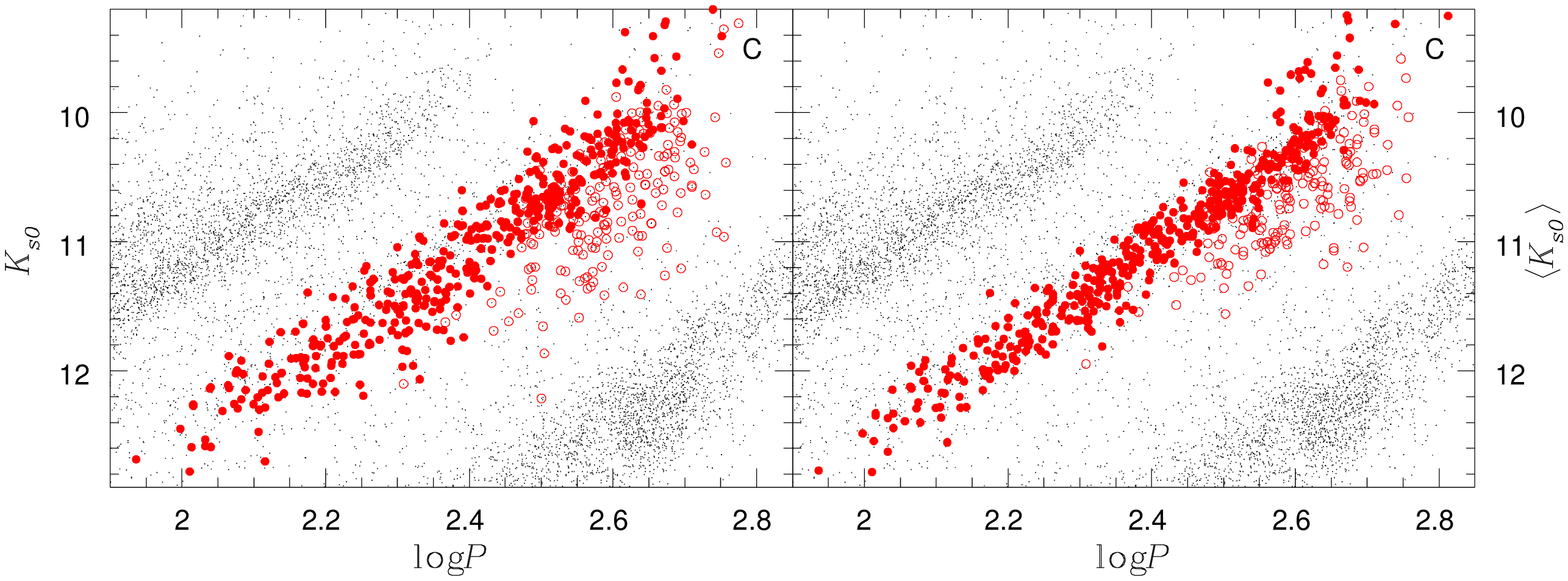}} 
\vspace{-4.0cm}
\FigCap{{\it Left panel}: the {\it PL} sequence C constructed using the 
single-epoch $K_{s0}$ magnitudes. {\it Right panel}: the same sequence
plotted with the mean magnitudes estimated using the method described in
the text. For clarity only the variables with {\it I}-band amplitudes
exceeding 0.5~mag are plotted. Empty circles indicate the stars with
$(J-K_s)>2.3$.}
\end{figure}

The efficiency of our mean $K_s$-band magnitudes estimation method is
demonstrated in Fig.~5. We show here the {\it PL} sequence C constructed
with the uncorrected single-epoch magnitudes, and the same diagram plotted
with the mean $K_s$ magnitudes estimated using our algorithm. For clarity
we present only the variables with {\it I}-band amplitudes exceeding
0.5~mag, for which the correction is the largest. The improvement of the
{\it PL} relation is clearly visible.

\Section{${\pmb K}_{\pmb s}$-band ${\pmb P}{\pmb L}$ Relations}
We used the mean dereddened $K_{s0}$ luminosities estimated with the
procedure described in the previous Section to construct the $PL(K)$
diagram (Fig.~6). The width of the sequences, in particular sequence C, are
significantly smaller in comparison with the diagram constructed with
uncorrected single-epoch measurements. In Fig.~6 the blue points represent
the O-rich stars and the red points mark the C-rich giants, both groups
discriminated in the period--$W_I$ diagram. Closer look at the $PL(K)$
diagram shows that both spectral types obey different relations between
$\log{P}$ and $K_{s0}$-band luminosity.

\begin{figure}[!t]
\centerline{\includegraphics[width=14cm]{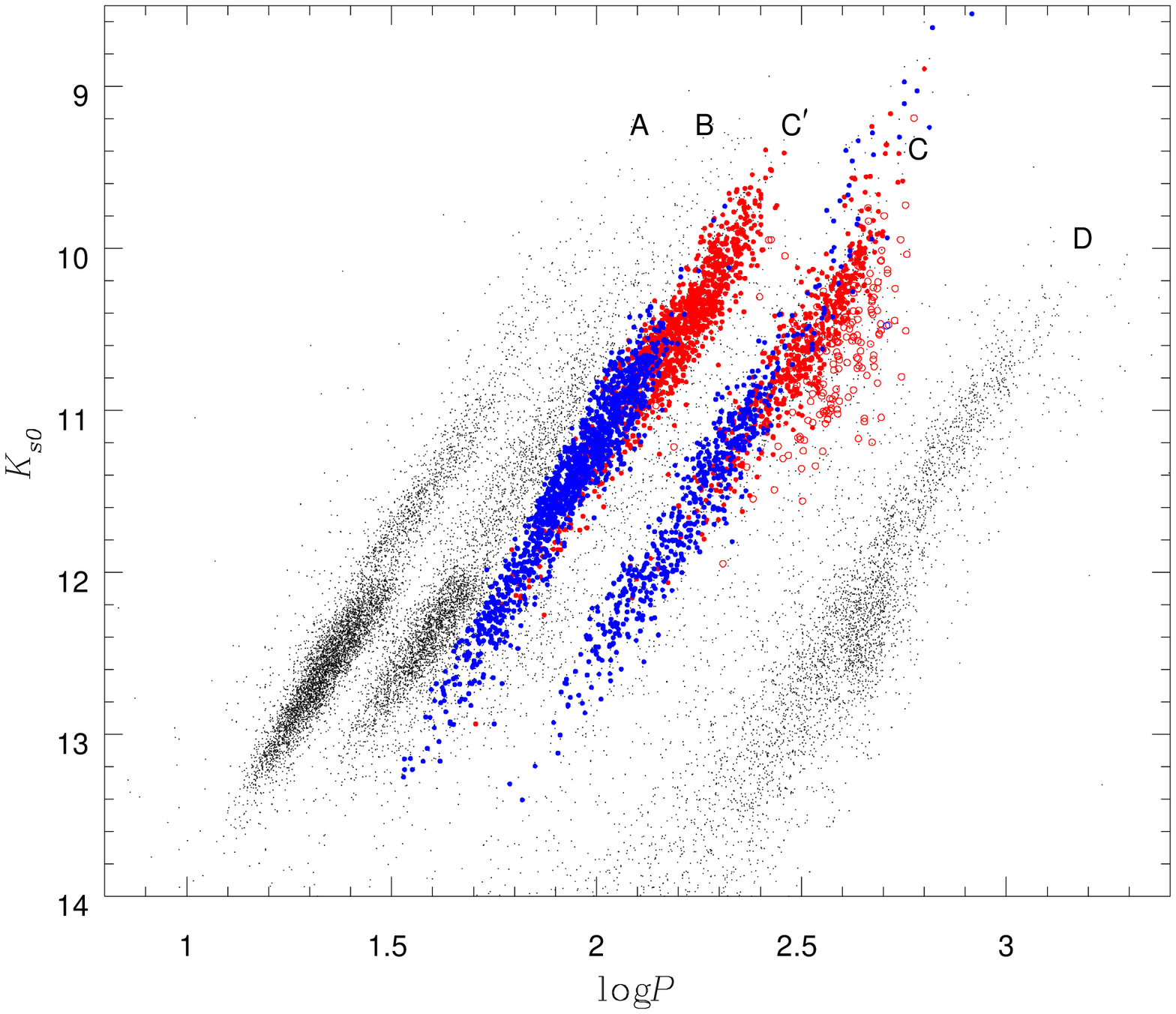}} 
\vspace{-6.0cm}
\FigCap{Period--$K_s$ diagram for Miras and SRVs in the LMC. The blue
points show the O-rich variables, while the red points mark the C-rich
stars. Empty circles indicate the stars with $(J-K_s)>2.3$~mag, \ie stars
surrounded by dust shells. The single-epoch $K_s$-band magnitudes of Miras
and SRVs were transformed to the mean magnitudes using the procedure
described in the text.}
\end{figure} 

Let us focus on some details seen in Fig.~6. Some brightest Mira-like
variables are located above the $PL(K)$ relationship C. Whitelock and Feast
(2000) suggested that these objects undergo the hot-bottom burning, but
their evolutionary status is unknown. Unfortunately, these stars are close
to the saturation limit of the OGLE-II photometry, and they are above this
limit in OGLE-III, thus only four years of observations are covered for
these stars. Other, shallower photometric surveys (\eg ASAS, Pojma{\'n}ski
2002) could be useful for studying these objects. Because these variables
do not follow the {\it PL} sequence of the remaining Mira-like variables,
we excluded them from fitting the linear {\it PL} relations.

The second feature that can be seen in the {\it PL} diagram is the fact
that the {\it PL} ridge of the C-rich Mira-like variables has considerably
larger dispersion than the O-rich sequence. Most of the outliers fall at
the right side of the mean {\it PL} relation, \ie they are fainter than the
typical C-Miras with the same periods. We checked that the vast majority of
the C-rich outliers are very red, with the $(J-K_s)_0>2.3$~mag. These
objects are marked with empty circles in Fig.~6. We suppose that these
stars are highly reddened due to a large amount of circumstellar material
around them. The same interpretation of the giants with $(J-K_s)>2.0$~mag
was presented by Nikolaev and Weinberg (2000). These object were also
excluded from the fit.

We used the least square method with $2.5\sigma$ clipping to fit the
$K_{s0}$-band {\it PL} relations separately for M- and C-type stars in the
sequences C (with the standard deviation in brackets):
$${\rm O-rich:~~}K_{s0}=-3.89(\pm0.05)(\log{P}-2.4)+10.923(\pm0.011)~~~(\sigma=0.14)$$
$${\rm C-rich:~~}K_{s0}=-3.71(\pm0.05)(\log{P}-2.4)+11.012(\pm0.008)~~~(\sigma=0.13)$$
and in the sequence C$'$:
$${\rm O-rich:~~}K_{s0}=-4.13(\pm0.03)(\log{P}-2.4)+9.551(\pm0.013)~~~(\sigma=0.13)$$
$${\rm C-rich:~~}K_{s0}=-3.76(\pm0.04)(\log{P}-2.4)+9.775(\pm0.010)~~~(\sigma=0.15)$$

It can be noticed that the {\it PL} relations of the O-rich and C-rich LPVs
have different slopes -- the C-type stars obey slightly shallower $PL(K)$
relations than the M-giants. The previous works (Feast 1984, Feast \etal
1989, Hughes and Wood 1990) suggested the same conclusions, but the
uncertainties of the slope determinations were much too large to confirm
this effect. Ita \etal (2004) used OGLE-II and SIRIUS data to show that the
slopes of O- and C-rich LPVs in the sequences C and C$'$ are different, but
the errors of their determinations were only two times smaller than the
difference of the slopes. Here, for the first time we show with the
reliability larger than $3\sigma$ that the slopes of the $PL(K)$ relations
for the O-rich LPVs are different than for the C-rich variables.

\Section{Discussion}
Now, it is possible to explain the existence of the additional two
sequences found in the $\log{P}$--$W_I$ diagram by Soszy{\'n}ski \etal
(2004ab): the sequence C$''$ spreading between the sequences C and D, and
D$'$ occupied the region of longer periods than the sequence D. Both ridges
are formed by C-type stars: sequence C$''$ is populated by Mira-like
variables and the sequence D$'$ by C-rich stars with the Long Secondary
Periods. The C-type SRVs from the sequence C$'$ overlap in the
$\log{P}$--$W_I$ diagram with the M-type Mira-like variables, so this
sequence has not been detected earlier.

A careful investigation of the {\it PL} diagrams gives us some clues
concerning the nature of the different {\it PL} relations in the
$\log{P}$--$K_s$ plane. The number of the obscured C-rich Miras (with
$(J-K_s)_0>2.3$~mag) which do not fit the $K_s$-band {\it PL} relation,
fall very close to the mean $\log{P}$--$W_{JK}$ relation. That can be
expected, because the Wesenheit index is the reddening independent
quantity.

The more surprising feature is that the C-rich and O-rich giants seem to
obey the same $\log{P}$--$W_{JK}$ relations in both, C and C$'$
sequences. There are two explanations of that fact. First, since the $J$
waveband spectral range of the C-rich AGB stars is strongly affected by the
CN and C$_2$ molecular bands (Cohen \etal 1981), the {\it J}-band mean
luminosities of these objects are on average fainter than the O-rich stars,
and the $(J-K_s)$ colors are larger. Thus, in the $\log{P}$--$W_{JK}$ plane
the C-type variables are located relatively higher, and by a coincidence
the period--$W_{JK}$ relations of both spectral types are the same.

The second explanation assumes that if the reddening free
$\log{P}$--$W_{JK}$ relations are the same for O-rich and C-rich giants,
than the difference between $K_s$-band {\it PL} relations of both classes
is only an effect of dust obscuration, on average larger in the C-type
Miras and SRVs. If the second interpretation is correct, our results
paradoxically confirm the idea that un-obscured C- and O-rich Miras fit the
same $PL(K)$ relation (Feast \etal 1989).

A completely different phenomenon must be responsible for the split of
sequences C and C$'$ in the period--$W_I$ plane. It is known that {\it
V}-band magnitudes of the M-type LPVs are strongly affected by the titanium
oxide (TiO) absorption (Smak 1964, Reid and Goldston 2002). Thus, O-rich
stars have larger $(V-I)$ colors than C-rich giants of same $K_s$-band
luminosity, and in consequence they lie higher (for the same periods) in
the period--$W_I$ diagram.

Using our method of discriminating the O- and C-rich AGB stars we can
derive the relative numbers of C- to M-type AGB stars (C/M ratio). The C/M
ratio is considered to be an indicator of the mean metallicity of a
stellar population as well as a tracer of the history of star formation
(Cioni \etal 2005).

\MakeTable{c@{\hspace{10pt}}c@{\hspace{10pt}}c@{\hspace{10pt}}c
@{\hspace{10pt}}}{12.5cm}{The C/M ratios for stars from the sequences C, 
C$'$ and D}
{\hline
\noalign{\vskip3pt}
 & Sequence C & Sequence C$'$ & Sequence D \\
\noalign{\vskip3pt}
\hline
\noalign{\vskip3pt}
All & 1.21 & 0.76 & -- \\
Above the TRGB & 1.64 & 0.92 & 0.20 \\
\noalign{\vskip3pt}
\hline
}

The C/M ratios for stars in the sequences C, C$'$, and D are shown in
Table~1. Because the sequence D undoubtedly contains a fraction of first
ascent giants, we present also the C/M values for stars above the TRGB
($K_s<12.05$~mag). It is clear that the largest relative number of C-rich
stars occurs in the Mira-like variables, it is intermediate for stars in
the sequence C$'$ and smallest for stars with the Long Secondary Periods. We
believe that this is an indicator of the evolutionary status of these
groups of AGB stars. The vast majority of variables in the sequence D have
secondary periods which falls within the four OSARG {\it PL} sequences
(Soszy{\'n}ski \etal 2004a). Thus, our results confirm the fact that the
longer the period of the LPVs, the more evolutionary advanced the star is.

It is worth to emphasize that, as can be noticed from Table~1, the C/M
ratio depends strongly on the method of the AGB stars selection. Different
relative numbers of O- and C-rich stars will be obtained when one considers
only Miras, Miras and SRVs, or all AGB stars. Likewise, different C/M ratio
will occur when one excludes the stars fainter than the TRGB (the bandpass
in which the TRGB is determined also matters) or considers all AGB
stars. Therefore, the comparison of the C/M in various environments can be
reliable only when the same criteria of the AGB stars selection are
fulfilled. This may be the source of large uncertainty in calibration of
the C/M \vs [Fe/H] relationship.

\Section{Summary}
Based on the OGLE-II and OGLE-III photometry supplemented by the 2MASS NIR
data we selected the sample of Miras and SRVs in the LMC. We showed that for
these objects the strip pattern of the {\it PL} distribution is more
complex than it was considered before. The sequences C, C$'$ and D split into
two separated ridges in the period--$W_I$ plane. This feature can be used
for the distinction between the O-rich and C-rich LPVs.

We applied the method of recovering mean $K_s$ magnitudes from single-phase
observations and complete {\it I}-band light curves. The scatter of the
$K_s$-band {\it PL} sequences, in particular sequence C, considerably
decreased in comparison with uncorrected magnitudes. We use these data to
show that the $K_s$-band period--luminosity relations are different for
O-rich and C-rich LPVs.

\vspace{5mm}
\Acknow{We are grateful to Prof.~Wojciech Dziembowski and
Prof.~Bohdan Paczy{\'n}ski for useful comments on this paper. The paper was
partly supported by the KBN grant 2P03D02124 to A.~Udalski. Partial support
to the OGLE project was provided with the NSF grant AST-0204908 and NASA
grant NAG5-12212 to B.~Paczy\'nski.

This publication makes use of data products from the Two Micron All Sky
Survey, which is a joint project of the University of Massachusetts and the
Infrared Processing and Analysis Center/California Institute of Technology,
funded by the National Aeronautics and Space Administration and the
National Science Foundation.

This paper is partly based on DENIS data obtained at the European Southern
Observatory.}

\end{document}